# Quantum and Relativistic corrections to Maxwell-Boltzmann ideal gas model from a Quantum Phase Space approach

**Rivo Herivola Manjakamanana Ravelonjato[1], Ravo Tokiniaina Ranaivoson[2], Raoelina Andriambololona[3], Roland Raboanary[4], Hanitriarivo Rakotoson[5], Naivo Rabesiranana[6]**

*manjakamanana@yahoo.fr[1], tokhiniaina@gmail.com[2], raoelina.andriambololona@gmail.com[3], r_raboanary@yahoo.fr[4], infotsara@gmail.com[5], rabesiranana@yahoo.fr[6]*

[1,2,3,5,6] Institut National des Sciences et Techniques Nucléaires (INSTN- Madagascar)
*BP 3907 Antananarivo 101, Madagascar, instn@moov.mg*

[4,5,6] Department of Physics, Faculty of Sciences – University of Antananarivo
*BP 566 Antananarivo 101, Madagascar*

*Abstract:* The quantum corrections related to the ideal gas model that are often considered are those which are related to the particles nature: bosons or fermions. These corrections lead respectively to the Bose-Einstein and Fermi-Dirac statistics. However, in this work, other kinds of corrections which are related to the quantum nature of phase space are considered. These corrections are introduced as improvement in the expression of the partition function of an ideal gas. Then corrected thermodynamics properties of the gas are deduced. Both the non-relativistic quantum and relativistic quantum cases are considered. It is shown that the corrections in the non-relativistic quantum case may be particularly useful to describe the deviation from classical behavior of a Maxwell-Boltzmann gas at low temperature and in confined space. These corrections can be considered as including the description of quantum size and shape effects. For the relativistic quantum case, the corrections could be relevant for confined space and when the thermal energy of each particle is comparable to their rest energy. The corrections appear mainly as modifications in the thermodynamic equation of state and in the expressions of the partition function and thermodynamic functions like entropy, internal energy, and free energy. Classical expressions are obtained as asymptotic limits.

# 1. Introduction

Since the advent of quantum physics and the theory of special relativity, various approaches have been considered to include quantum and relativistic corrections in the ideal gas model or more generally in statistical mechanics and thermodynamic [1-21]. In this work, we consider the study on new kind of corrections for the Maxwell-Boltzmann ideal gas model which can be considered as related with the quantum nature of phase space. The approach uses results from the references [22-24].

The Maxwell-Boltzmann ideal gas model is the classical simplest model that is used to describe gas behavior and to illustrate basic relations between thermodynamics and statistical mechanics. In this model, gas particles are approximately considered to be free punctual particles and their individual energy is equal to their kinetic energy. In the framework of classical statistical mechanics, the partition function of each particle, i.e. the single-particle partition function is given by the relation:

$$\zeta_C = \frac{1}{h^3}\int e^{-\beta\frac{\vec{p}^2}{2m}}d^3\vec{x}d^3\vec{p} = \frac{V}{h^3}\left(\frac{2\pi m}{\beta}\right)^{3/2} = \frac{V}{h^3}(2\pi m k_B T)^{3/2} = \frac{V}{(\lambda_{th})^3} \qquad (1)$$

in which:
- $h$ is the Planck constant and $k_B$ is the Boltzmann constant
- $\beta = \frac{1}{k_B T}$ with $T$ and $V$ being respectively the temperature and volume of the gas.
- $m$ is the mass of a gas particle, $\vec{x}$ is its position vector and $\vec{p}$ its linear momentum vector.
- $\lambda_{th} = \frac{h}{(2\pi m k_B T)^{1/2}}$ is the thermal de Broglie wavelength

The integration which is considered in this expression is performed over the phase space domain $\{(\vec{x},\vec{p})\}$ which corresponds to the set of the possible classical microstates of the particle. The Planck constant $h$ appears here because it is related to the fact that the phase space hypervolume of a microstate is finite (not equal to zero) [24-25].

For an ideal gas composed of $N$ particles, the canonical classical partition function $\mathcal{Z}_C$ is approximately given by the relation (particles are considered as indiscernible)

$$\mathcal{Z}_C = \frac{1}{N!}(\zeta_C)^N = \frac{V^N}{h^{3N}N!}(2\pi m k_B T)^{\frac{3N}{2}} = \frac{V^N}{N!(\lambda_{th})^{3N}} \qquad (2)$$

Any thermodynamic properties of the gas can be deduced from the partition function $\mathcal{Z}_C$. We have for instance, in conventional classical thermodynamics, for the expressions of the classical internal energy $U$, the entropy $S$, the free energy $F$ and the pressure $P$ [2-4]:

$$\begin{cases} U = -\frac{1}{\mathcal{Z}_C}\left(\frac{\partial \mathcal{Z}_C}{\partial \beta}\right)_{V,N} = -\left(\frac{\partial \ln \mathcal{Z}_C}{\partial \beta}\right)_{V,N} = \frac{3}{2}Nk_B T \\ S = k_B[\ln(\mathcal{Z}_C) + \beta U] \cong Nk_B\left[\frac{5}{2} + \ln\left(\frac{V}{N(\lambda_{th})^3}\right)\right] \\ F = U - TS = -k_B T \ln(\mathcal{Z}_C) \cong Nk_B T\left[\ln\left(\frac{N(\lambda_{th})^3}{V}\right) - 1\right] \\ P = -\left(\frac{\partial F}{\partial V}\right)_{T,N} = k_B T\left(\frac{\partial \ln \mathcal{Z}_C}{\partial V}\right)_{T,N} = \frac{N}{V}k_B T \end{cases} \qquad (3)$$

The use of the symbol "≅" in the explicit expressions of the entropy $S$ and the free energy $F$ correspond to the fact that the Stirling's approximation $ln(N!) \cong NlnN - N$ was used. The explicit expression of the entropy corresponds to the Sackur-Tetrode equation. The well-known thermodynamic equation of state of a Maxwell-Boltzmann ideal gas can be deduced easily from the expression of the pressure $P$

$$PV = Nk_BT \qquad (4)$$

The quantum corrections related to the ideal gas model often considered are those related to the nature of the particles: bosons or fermions. These corrections lead respectively to the Bose-Einstein and Fermi-Dirac statistics [2-5]. However, the corrections related to quantum size and shape effects may become important in some interesting cases like in the field of nanoscience and nanotechnology [18-19], [26-28]. In [16], for instance, the quantum size effect corrections to the ideal gas model is considered and deduced as a consequence of the quantization of the energy of a particle in a box. In fact, for a particle confined in a parallelepiped with volume $V$ equal to $L_1L_2L_3$, it can be deduced from the Schrodinger equation that the expression of the energy is

$$\varepsilon_{n_1,n_2,n_3} = \frac{\pi^2\hbar^2}{2m}\left[\left(\frac{n_1}{L_1}\right)^2 + \left(\frac{n_2}{L_2}\right)^2 + \left(\frac{n_3}{L_3}\right)^2\right] \qquad (5)$$

in which $n_1, n_2$ and $n_3$ are integers, $\hbar$ is the reduced Planck and $m$ is the mass of the particle. Then, instead of the relation (1) one should have for the single-particle partition function

$$\zeta = \sum_{n_1=0}^{+\infty}\sum_{n_2=0}^{+\infty}\sum_{n_3=0}^{+\infty} e^{-\beta\frac{\pi^2\hbar^2}{2m}\left[\left(\frac{n_1}{L_1}\right)^2 + \left(\frac{n_2}{L_2}\right)^2 + \left(\frac{n_3}{L_3}\right)^2\right]} \qquad (6)$$

It is very difficult, if not impossible, to find an analytical compact form for the exact result of the summation (6). At macroscopic level (when $L_1, L_2$ and $L_3$ is large compared to $\pi\hbar\sqrt{\frac{\beta}{2m}}$) the summation (6) is approximately equal to the integral (1). But at microscopic scale, the difference may become significant. In the reference [16], the quantum size effect correction is based on the relation (6) but an approximate evaluation of this summation was used.

In this work, we use an alternative method, based on the consideration of the quantum nature of phase space to introduce corrections related to the quantum size and shape effects. Relativistic corrections are studied too. Our approach is based on the concepts of quantum phase space and phase space representation of quantum mechanics that were considered in the references [22-24]. These concepts are described briefly in the section 2. The expressions of the single-particle Hamiltonian operators are also determined in this section both for the non-relativistic and the relativistic cases. The study of the quantum corrections are performed in the section 3 and the relativistic quantum corrections are considered in the section 4. Discussions are given in the section 5 before the conclusion section. The used notation system is inspired from the reference [29]. Boldfaced letters are used to represent quantum operators. The abbreviations R-Q, NR-Q, R-NQ, NR-NQ stand respectively for Relativistic Quantum, Non-Relativistic Quantum, Relativistic Non-Quantum, Non-Relativistic and Non-Quantum. The notation $tr$ and $det$ correspond to the trace and determinant of an operator or a matrix.

## 2. Quantum phase space and single-particle Hamiltonian operator

### 2.1. Quantum phase space and phase space representation

In classical physics, phase space is intuitively defined as the set of possible values of the spatial coordinates and the components of the linear momentum vector of the particle. The extension of this definition in the context of quantum physics is not straightforward because of the uncertainty relations [7, 22-24]. Various attempts have already been made to tackle the problem of phase space in quantum physics. One of the most famous corresponds to what is called phase space formulation of quantum mechanics [30-31]. This formulation has been introduced and developed firstly in the references [1, 32-34].This approach deals with a formulation of quantum mechanics but the phase space which is considered is a classical one. Another approach has been considered and developed through the works [22-24] in which the concept of "quantum phase space" is introduced. For a monodimensional case, this quantum phase space was defined as the set $\{(\langle x \rangle, \langle p \rangle)\}$ of the possible mean values $\langle x \rangle$ and $\langle p \rangle$ of the coordinate operator $x$ and momentum operator $p$ corresponding to quantum states denoted $|\langle z \rangle\rangle$ which are themselves eigenstates of the operator $z$ defined by the relation [22-24]

$$\boldsymbol{z} = \boldsymbol{p} - \frac{2i}{\hbar}\mathcal{B}\boldsymbol{x} \tag{7}$$

in which $\mathcal{B}$ is the statistical variance of the momentum operator corresponding to the state $|\langle z \rangle\rangle$ itself. The corresponding eigenvalue equation is

$$\boldsymbol{z}|\langle z \rangle\rangle = \langle z \rangle|\langle z \rangle\rangle = (\langle p \rangle - \frac{2i}{\hbar}\mathcal{B}\langle x \rangle)|\langle z \rangle\rangle \tag{8}$$

and the wavefunction in coordinate representation corresponding to a state $|\langle z \rangle\rangle$ is

$$\langle x|\langle z \rangle\rangle = (\frac{2\mathcal{B}}{\pi\hbar^2})^{1/4} e^{-\frac{\mathcal{B}}{\hbar^2}(x-\langle x \rangle)^2 + \frac{i}{\hbar}\langle p \rangle x} = (\frac{1}{2\pi\mathcal{A}})^{1/4} e^{-\frac{1}{4\mathcal{A}}(x-\langle x \rangle)^2 + \frac{i}{\hbar}\langle p \rangle x} \tag{9}$$

in which $\mathcal{A}$ is the statistical variance of the coordinate operator (corresponding to a state $|\langle z \rangle\rangle$). The relation between $\mathcal{A}$ and $\mathcal{B}$ corresponds to a saturation of the uncertainty relation

$$\mathcal{A}\mathcal{B} = \frac{\hbar^2}{4} \tag{10}$$

It is possible to define a phase space representation of quantum mechanics using the set $\{|\langle z \rangle\rangle\}$ of the eigenstates $|\langle z \rangle\rangle$ of the operator $\boldsymbol{z}$ [23-24].

### 2.2. Hamiltonian operator for a non-relativistic particle

It was shown in [23] that the expression of the Hamiltonian operator $\boldsymbol{H}_Q$ of a nonrelativistic free particle, for a monodimensional motion, compatible with the uncertainty principle and the concept of quantum phase space is

$$\boldsymbol{H}_Q = \frac{(\langle p \rangle)^2}{2m} + \frac{\mathcal{B}}{m}(\boldsymbol{z}^\dagger\boldsymbol{z} + \boldsymbol{z}\boldsymbol{z}^\dagger) = \frac{(\langle p \rangle)^2}{2m} + \frac{\mathcal{B}}{m}(2\boldsymbol{z}^\dagger\boldsymbol{z} + 1) \tag{11}$$

in which $\boldsymbol{z}$ and $\boldsymbol{z}^\dagger$ are operators related with the operator $\boldsymbol{z}$ in (7) by the relation

$$\begin{cases} \boldsymbol{z} = i\dfrac{(\boldsymbol{z} - \langle z \rangle)}{2\sqrt{\mathcal{B}}} = \dfrac{i}{2\sqrt{\mathcal{B}}}\left[(\boldsymbol{p} - \langle p \rangle) - 2\dfrac{i}{\hbar}\mathcal{B}(\boldsymbol{x} - \langle x \rangle)\right] \\ \boldsymbol{z}^\dagger = -i\dfrac{(\boldsymbol{z}^\dagger - \langle z \rangle^*)}{2\sqrt{\mathcal{B}}} = \dfrac{-i}{2\sqrt{\mathcal{B}}}\left[(\boldsymbol{p} - \langle p \rangle) + 2\dfrac{i}{\hbar}\mathcal{B}(\boldsymbol{x} - \langle x \rangle)\right] \end{cases} \quad (12)$$

They verify the commutation relation (which characterizes ladder operators)

$$[\boldsymbol{z}, \boldsymbol{z}^\dagger] = 1 \quad (13)$$

In the reference [23], the eigenstates of the Hamiltonian operator in (11) are the quantum states denoted $|n, \langle z \rangle\rangle$ which can be deduced from a state $|\langle z \rangle\rangle$ by the relation

$$|n, \langle z \rangle\rangle = \dfrac{(\boldsymbol{z}^\dagger)^n}{\sqrt{n!}}|\langle z \rangle\rangle \quad (14)$$

with $n$ a positive integer. The corresponding eigenvalue equation is

$$\boldsymbol{H}_Q|n, \langle z \rangle\rangle = \varepsilon_n|n, \langle z \rangle\rangle = \left[\dfrac{\langle \vec{p} \rangle^2}{2m} + (2n+1)\dfrac{\mathcal{B}}{m}\right]|n, \langle z \rangle\rangle \quad (15)$$

For a fixed value of $\langle z \rangle$, the set $\{|n, \langle z \rangle\rangle\}$ is an orthonormal basis of the state space of the particle. The possible values $\varepsilon_n$ of the energy of the particle are given by the relation

$$\varepsilon_n = \dfrac{\langle \vec{p} \rangle^2}{2m} + (2n+1)\dfrac{\mathcal{B}}{m} \quad (16)$$

$\varepsilon_n$ consists of two parts: the first one is a classical-like mean kinetic energy and the second one is related to the quantum uncertainty relation.

We may define a "mean rest frame" of the particle in which one has $\langle \vec{p} \rangle = \vec{0}$. In this frame, we have for the Hamiltonian operator and its eigenvalues

$$\begin{cases} \boldsymbol{H}_Q = (2\boldsymbol{z}^\dagger\boldsymbol{z} + 1)\dfrac{\mathcal{B}}{m} \\ \varepsilon_n = (2n+1)\dfrac{\mathcal{B}}{m} \end{cases} \quad (17)$$

For a tridimensional case, the generalization of the relation (17) is [19]:

$$\begin{cases} \boldsymbol{H}_Q = \sum_{l=1}^{3}(2\boldsymbol{z}_l^\dagger\boldsymbol{z}_l + 1)\dfrac{\mathcal{B}_{ll}}{m} \\ \varepsilon_n = \sum_{l=1}^{3}(2n_l + 1)\dfrac{\mathcal{B}_{ll}}{m} \end{cases} \quad (18)$$

in which the index $n$ of $\varepsilon_n$ corresponds to the triplet of integer numbers $n = (n_1, n_2, n_3)$.

### 2.3. Hamiltonian operator for a relativistic quantum particle

In relativistic and non-quantum (R-NQ) mechanics, the energy $\varepsilon$ of a particle with mass $m$ and linear momentum $\vec{p}$ is given by the relation ($c$ being the speed of light):

$$\varepsilon^2 = m^2c^4 + \vec{p}^2c^2 \qquad (19)$$

It may be deduced from the relations (11) and (19) that the Hamiltonian operator $\boldsymbol{H}_{QR}$ for a relativistic quantum particle should satisfies the following relation

$$(\boldsymbol{H}_{RQ})^2 = m^2c^4 + \langle\vec{p}\rangle^2 c^2 + \frac{2\mathcal{B}}{m}(2\boldsymbol{z}^\dagger\boldsymbol{z} + 1) \qquad (20)$$

The Hamiltonian operator $\boldsymbol{H}_Q$ in the relation (11) associated to the NR-Q case can be considered as corresponding to the kinetic energy part (including the quantum fluctuation part) of the RQ Hamiltonian operator $\boldsymbol{H}_{RQ}$ in (20) in the NR-Q limit (when the kinetic energy is small compared to the rest energy $mc^2$). The eigenstates of the operator $(\boldsymbol{H}_{RQ})^2$ are also the states $|n, \langle z\rangle\rangle$ corresponding to the relation (14). The corresponding eigenvalue equation is

$$(\boldsymbol{H}_{RQ})^2|n,\langle z\rangle\rangle = (\varepsilon_n)^2|n,\langle z\rangle\rangle = \left[m^2c^4 + \langle\vec{p}\rangle^2c^2 + 2(2n+1)\frac{\mathcal{B}c^2}{m}\right]|n,\langle z\rangle\rangle \qquad (21)$$

From this relation (21), the possible values $(\varepsilon_n)^2$ of the square of the energy of the particle are given by the relation

$$(\varepsilon_n)^2 = m^2c^4 + \langle\vec{p}\rangle^2c^2 + 2(2n+1)\mathcal{B}c^2 \qquad (22)$$

The non-relativistic energies given by the relation (16) corresponds to the kinetic part of the relativistic energies associated to the relation (22) in the non-relativistic limit (when this kinetic part is small compared to the rest energy $mc^2$).

Using the "mean rest frame" of the particle (in which $\langle\vec{p}\rangle = \vec{0}$), the relativistic generalization corresponding to the tridimensional relations in (18) are

$$\begin{cases} (\boldsymbol{H}_{RQ})^2 = m^2c^4 + 2\sum_{l=1}^{3}(2\boldsymbol{z}_l^\dagger\boldsymbol{z}_l + 1)\mathcal{B}_{ll}c^2 \\ (\varepsilon_n)^2 = m^2c^4 + 2\sum_{l=1}^{3}(2n_l + 1)\mathcal{B}_{ll}c^2 \end{cases} \qquad (23)$$

**Remark**: It should be pointed out that the relations in (18) and (23) correspond to what was called 'mean rest frame' of the particle in which the mean value of the linear momentum of the particle is equal to zero: $\langle\vec{p}\rangle = \vec{0}$. This choice is motivated by the fact that for particles composing an ideal gas the relation $\langle\vec{p}\rangle = \vec{0}$ is satisfied. It follows that the frame in which the container which contains the gas is at rest can be considered as the 'mean rest frame' of the particles.

## 3. Quantum corrections to the ideal gas model

### 3.1. Single-particle non-relativistic quantum partition function

Using results from the reference [19], it may be deduced that the density operator $\boldsymbol{\rho}$ of a non-relativistic particle at equilibrium with a heat bath of temperature $T$ is given by the relation (quantum canonical distribution)

$$\boldsymbol{\rho}_Q = \frac{1}{\zeta_Q} e^{-\beta H_Q} \qquad (24)$$

with $\beta = \frac{1}{k_B T}$ (the same as in the relation (1)). $\zeta_Q$ is the non-relativistic quantum single-particle partition function and $\boldsymbol{H}_Q$ the non-relativistic Hamiltonian given in the relation (18). The expression of $\zeta_Q$ is

$$\zeta_Q = tr(e^{-\beta H_Q}) = \sum_{n_1,n_2,n_3} e^{-\beta \sum_{l=1}^{3}(2n_l+1)\frac{\mathcal{B}_{ll}}{m}} = \frac{1}{8 sh\left(\frac{\beta}{m}\mathcal{B}_{11}\right) sh\left(\frac{\beta}{m}\mathcal{B}_{22}\right) sh\left(\frac{\beta}{m}\mathcal{B}_{33}\right)} \qquad (25)$$

Another way to calculate the partition function $\zeta_Q$ in (25) is to use the basis $\{|\langle\vec{z}\rangle\rangle\}$ composed by the eigenstates of the operator $\vec{\mathbf{z}}$ [19]:

$$\zeta_Q = Tr(e^{-\beta H_Q}) = \int \langle\langle\vec{z}\rangle| e^{-\beta H_Q} |\langle\vec{z}\rangle\rangle \frac{d^3\langle\vec{p}\rangle d^3\langle\vec{x}\rangle}{h^3} \qquad (26)$$

At the classical limit $\frac{\beta}{m}\mathcal{B}_{ll} \ll 1$:

- On one hand, one has $sh(\frac{\beta}{m}\mathcal{B}_{ii}) \cong \frac{\beta}{m}\mathcal{B}_{ii}$ at first order, then (25) becomes

$$\zeta_Q = \frac{1}{8 sh\left(\frac{\beta}{m}\mathcal{B}_{11}\right) sh\left(\frac{\beta}{m}\mathcal{B}_{22}\right) sh\left(\frac{\beta}{m}\mathcal{B}_{33}\right)} \cong \frac{m^3}{8\beta^3 \mathcal{B}_{11}\mathcal{B}_{22}\mathcal{B}_{33}} \qquad (27)$$

- On the other hand, we obtain for the expression (26) the approximation

$$\zeta_Q = \int \langle\langle\vec{z}\rangle| e^{-\beta H_Q} |\langle\vec{z}\rangle\rangle \frac{d^3\langle\vec{p}\rangle d^3\langle\vec{x}\rangle}{h^3} \cong \int e^{-\beta \frac{\langle\vec{p}\rangle^2}{2m}} \frac{d^3\langle\vec{p}\rangle d^3\langle\vec{x}\rangle}{h^3} = \frac{V}{(\lambda_{th})^3} = \zeta_C \qquad (28)$$

in which $\zeta_C$ is the classical partition function in the relation (1). The identification between the relations (27) and (28) permits to determine the relations between the momentum statistical variances $\mathcal{B}_{ll}$ and the thermodynamics variables as in [19]. One find for instance for a parallelepipedic volume $V$ equal to $L_1 L_2 L_3$

$$\begin{cases} \mathcal{B}_{11} = \dfrac{\hbar^2}{2L_1 \lambda_{th}} = \dfrac{\hbar}{2L_1}\sqrt{\dfrac{m}{2\pi\beta}} = \dfrac{\hbar}{2L_1}\sqrt{\dfrac{mk_BT}{2\pi}} \\[6pt] \mathcal{B}_{22} = \dfrac{\hbar^2}{2L_2 \lambda_{th}} = \dfrac{\hbar}{2L_2}\sqrt{\dfrac{m}{2\pi\beta}} = \dfrac{\hbar}{2L_2}\sqrt{\dfrac{mk_BT}{2\pi}} \\[6pt] \mathcal{B}_{33} = \dfrac{\hbar^2}{2L_3 \lambda_{th}} = \dfrac{\hbar}{2L_3}\sqrt{\dfrac{m}{2\pi\beta}} = \dfrac{\hbar}{2L_3}\sqrt{\dfrac{mk_BT}{2\pi}} \end{cases} \quad (29)$$

### 3.2. Non-relativistic Quantum partition function for an ideal gas

An ideal gas can be considered as a set of $N$ indiscernible particles. The expression of the corresponding partition function (which do not take into account the nature of the particles: bosons or fermions) is approximately given by the relation

$$\mathcal{Z}_Q = \frac{(\zeta_Q)^N}{N!} \qquad (30)$$

in which $\zeta_Q$ is the one particle partition function which is given by the relation (25). From the relations (25) and (30), one can deduces

$$\mathcal{Z}_Q = \frac{1}{8^N N! \left[ sh\left(\frac{\beta}{m}\mathcal{B}_{11}\right) sh\left(\frac{\beta}{m}\mathcal{B}_{22}\right) sh\left(\frac{\beta}{m}\mathcal{B}_{33}\right) \right]^N} = \frac{1}{8^N N! \{det[sh(\frac{\beta}{m}[\mathcal{B}])]\}^N} \qquad (31)$$

in which $[\mathcal{B}]$ is the matrix corresponding to the momentum statistical variances $\mathcal{B}_{ll}$ and $det$ is referring to matrix determinant.

### 3.3. Thermodynamic properties

The thermodynamic properties of the gas can be deduced from the partition function $\mathcal{Z}_Q$ in the relation (31). One obtains respectively for the internal energy $U_Q$, the entropy $S_Q$, the free energy $F_Q$

$$\begin{cases} U_Q = \langle \boldsymbol{H_Q} \rangle = \dfrac{1}{\mathcal{Z}_Q}\left(\dfrac{\partial \mathcal{Z}_Q}{\partial \beta}\right) = \dfrac{1}{2}Nk_BT\left[Tr\left(\dfrac{\beta}{m}[\mathcal{B}]\,coth(\dfrac{\beta}{m}[\mathcal{B}])\right)\right] \\[6pt] S_Q = k_B[\ln(\mathcal{Z}_Q) + \beta U_Q] = -k_B \ln\left[8^N N!\{det[sh(\dfrac{\beta}{m}[\mathcal{B}])]\}^N\right] + \dfrac{U_Q}{T} \\[6pt] F_Q = U_Q - TS_Q = -k_BT\ln\left[8^N N!\{det[sh(\dfrac{\beta}{m}[\mathcal{B}])]\}^N\right] \end{cases} \qquad (32)$$

For the pressure, we do not have a scalar like in the classical case. But we obtain a pressure matrix (analogous to the opposite of a stress tensor) $[P]$ which is related to the matrix $[\mathcal{B}]$ of the momentum statistical variances by the relation

$$[P] = \frac{N}{V}k_BT\frac{\beta}{m}[\mathcal{B}]coth\left(\frac{\beta}{m}[\mathcal{B}]\right) \qquad (33)$$

Explicitly, on has for instance for a parallelepiped volume $V$ equal to $L_1 L_2 L_3$

$$\begin{cases} P_{11} = -\dfrac{1}{L_2 L_3}\left(\dfrac{\partial F}{\partial L_1}\right)_{L_2,L_3,T,N} = \dfrac{Nk_B T}{V}\dfrac{\beta}{m}\mathcal{B}_{11}\coth\left(\dfrac{\beta}{m}\mathcal{B}_{11}\right) \\ P_{22} = -\dfrac{1}{L_1 L_3}\left(\dfrac{\partial F}{\partial L_2}\right)_{L_1,L_3,T,N} = \dfrac{Nk_B T}{V}\dfrac{\beta}{m}\mathcal{B}_{22}\coth\left(\dfrac{\beta}{m}\mathcal{B}_{22}\right) \\ P_{33} = -\dfrac{1}{L_1 L_2}\left(\dfrac{\partial F}{\partial L_3}\right)_{L_1,L_2,T,N} = \dfrac{Nk_B T}{V}\dfrac{\beta}{m}\mathcal{B}_{33}\coth\left(\dfrac{\beta}{m}\mathcal{B}_{33}\right) \end{cases} \quad (34)$$

The thermodynamic equation of state can be deduced easily from the relation (33)

$$[P]V = Nk_B T \frac{\beta}{m}[\mathcal{B}]\coth\left(\frac{\beta}{m}[\mathcal{B}]\right) \quad (35)$$

The relations in (3) and (4) can be considered as asymptotic limits of the relations in (32), (33) and (35) in the classical limits $\frac{\beta}{m}\mathcal{B}_{ll} \ll 1$. Taking account of the relations in (29) between the $\mathcal{B}_{ll}$ and the thermodynamic variables, these classical limits correspond to high temperature and large volume.

## 4. Relativistic and Quantum corrections to the ideal gas model
### 4.1. Single-particle partition function and thermodynamic properties

Given the relations (18), (23) and (24), one may suppose that the relativistic generalization of the relation (24) should be

$$\boldsymbol{\rho}_{RQ} = \frac{1}{\zeta_{RQ}} e^{-\beta_R (\boldsymbol{H}_{RQ})^2} \quad (36)$$

in which $\beta_R$ is a factor which will be determined in the relation (50) below. $\zeta_{RQ}$ is the relativistic quantum single-particle partition function and $(\boldsymbol{H}_{RQ})^2$ is given by the relation (23). It can be deduced from this relation (23) that the expression of $\zeta_R$ is

$$\zeta_{RQ} = tr\left(e^{-\beta_R (\boldsymbol{H}_R)^2}\right) = \sum_{n_1,n_2,n_3} e^{-\beta_R \sum_{l=1}^{3}[m^2 c^4 + 2\sum_{l=1}^{3}(2n_l+1)\frac{\mathcal{B}_{ll} c^2}{m}]} \quad (37a)$$

$$= \frac{e^{-\beta_R m^2 c^4}}{8\,sh(2\beta_R c^2 \mathcal{B}_{11})\,sh(2\beta_R c^2 \mathcal{B}_{22})\,sh(2\beta_R c^2 \mathcal{B}_{33})} = \frac{e^{-\beta_R m^2 c^4}}{8\,det[sh(2\beta_R c^2 [\mathcal{B}])]} \quad (37b)$$

Like in the non-relativistic case, another way to calculate the partition function $\zeta_R$ in (37) is to use the basis $\{|\langle \vec{z}\rangle\rangle\}$ composed by the eigenstates of the operator $\vec{\boldsymbol{z}}$

$$\zeta_{RQ} = Tr\left(e^{-\beta_R (\boldsymbol{H}_R)^2}\right) = \int \langle\langle \vec{z}\rangle| e^{-\beta_R (\boldsymbol{H}_{RQ})^2} |\langle \vec{z}\rangle\rangle \frac{d^3\langle\vec{p}\rangle d^3\langle\vec{x}\rangle}{h^3} \quad (38)$$

In the R-NQ limit $(2\beta_R c^2 \mathcal{B}_{ll}) \ll 1$ :

- On one hand, one has $sh(2\beta_R c^2 \mathcal{B}_{ll}) \cong 2\beta_R c^2 \mathcal{B}_{ll}$ at first order, then (37b) becomes

$$\zeta_{RQ} \cong \frac{e^{-\beta_R m^2 c^4}}{64(\beta_R)^3 c^6 \mathcal{B}_{11} \mathcal{B}_{22} \mathcal{B}_{33}} \qquad (39)$$

- On the other hand, we obtain for the expression (38) the approximation

$$\zeta_{RQ} \cong \int e^{-\beta_R (m^2 c^4 + \langle \vec{p} \rangle^2 c^2)} \frac{d^3 \langle \vec{p} \rangle d^3 \langle \vec{x} \rangle}{h^3} = \frac{V}{h^3} \left(\frac{\pi}{\beta_R c^2}\right)^{\frac{3}{2}} e^{-\beta_R m^2 c^4} \qquad (40)$$

The identification between the relations (38) and (39) permits to determine the relations between the momentum statistical variances $\mathcal{B}_{ll}$, the size and shape of the volume $V$ and $\beta_R$. One find for instance for a parallelepipedic volume $V$ equal to $L_1 L_2 L_3$

$$\begin{cases} \mathcal{B}_{11} = \dfrac{h}{4cL_1 \sqrt{\pi \beta_R}} \\ \mathcal{B}_{22} = \dfrac{h}{4cL_2 \sqrt{\pi \beta_R}} \\ \mathcal{B}_{33} = \dfrac{h}{4cL_3 \sqrt{\pi \beta_R}} \end{cases} \qquad (41)$$

It is to be remarked that the approximate expression of $\zeta_{RQ}$ in the non-quantum (but relativistic) limit (40) should be taken as the R-NQ single-particle partition function. Let us denote it $\zeta_R$:

$$\zeta_R = \frac{V}{h^3 c^3} \left(\frac{\pi}{\beta_R}\right)^{3/2} e^{-\beta_R m^2 c^4} \qquad (42)$$

From the expression (36) of the density operator, one may suppose that the relativistic quantum internal energy $u_{RQ}$ of the particle should satisfy the relation

$$(u_{RQ})^2 = \langle (\boldsymbol{H}_{RQ})^2 \rangle = tr[\boldsymbol{\rho}_{RQ} (\boldsymbol{H}_{RQ})^2] \qquad (43)$$

Explicitly, we have

$$(u_{RQ})^2 = \frac{1}{\zeta_{RQ}} \sum_{n_1, n_2, n_3} [m^2 c^4 + 2 \sum_{l=1}^{3} (2n_l + 1) \frac{\mathcal{B}_{ll} c^2}{m}] e^{-\beta_R \sum_{l=1}^{3} [m^2 c^4 + 2 \sum_{l=1}^{3} (2n_l + 1) \frac{\mathcal{B}_{ll} c^2}{m}]} \qquad (44)$$

From the relations (37a) and (44), between the partition function $\zeta_{RQ}$ and $(u_{RQ})^2$ we obtain the relation

$$(u_{RQ})^2 = -\frac{1}{\zeta_{RQ}} \frac{\partial \zeta_{RQ}}{\partial \beta_R} = -\frac{\partial ln(\zeta_{RQ})}{\partial \beta_R} \qquad (45)$$

and for the relativistic quantum entropy (Von-Neumann entropy):

$$s_{RQ} = k_B ln(\zeta_{RQ}) + k_B \beta_R (u_{RQ})^2 \tag{46}$$

From the relation (37b), (45) and (46), one can deduce

$$\begin{cases} (u_{RQ})^2 = m^2 c^4 + tr(c^2[\mathcal{B}] \coth(2\beta_R c^2[\mathcal{B}])) \\ s_{RQ} = k_B[\beta_R tr(c^2[\mathcal{B}] \coth(2\beta_R c^2[\mathcal{B}])) - ln\{det[sh(2\beta_R c^2[\mathcal{B}])]\}] \end{cases} \tag{47}$$

In the R-NQ limit $(2\beta_R c^2 \mathcal{B}_{ll}) \ll 1$, one can deduce from the relation (42) or the relations in (47) the expressions of the R-NQ internal energy $u_R$ and entropy $s_R$ of the particle:

$$\begin{cases} (u_R)^2 = -\dfrac{\partial \ln(\zeta_R)}{\partial \beta_R} = m^2 c^4 + \dfrac{3}{2\beta_R} \\ s_R = k_B ln(\zeta_R) + k_B \beta_R (u_R)^2 = \dfrac{3}{2} k_B + k_B ln[\dfrac{V}{h^3 c^3} \left(\dfrac{2\pi[(u_R)^2 - m^2 c^4)]}{3}\right)^{3/2}] \end{cases} \tag{48}$$

From thermodynamic, we have also the relation

$$\left(\dfrac{\partial s_R}{\partial u_R}\right)_V = \dfrac{1}{T} \Leftrightarrow \dfrac{3 k_B u_R}{(u_R)^2 - m^2 c^4} = \dfrac{1}{T} \tag{49}$$

The explicit expressions of $\beta_R, u_R$ and $s_R$ can be deduced from the relations (48) and (49)

$$\begin{cases} \beta_R = \dfrac{1}{2 k_B T u_R} = \dfrac{\sqrt{9(k_B T)^2 + 4m^2 c^4} - 3 k_B T}{4 m^2 c^4 k_B T} \\ u_R = \dfrac{1}{2 k_B T \beta_R} = \dfrac{3 k_B T + \sqrt{9(k_B T)^2 + 4m^2 c^4}}{2} \\ s_R = \dfrac{3}{2} k_B + k_B ln\left[\dfrac{V}{h^3 c^3} \left(\dfrac{\pi}{\beta_R}\right)^{\frac{3}{2}}\right] \end{cases} \tag{50}$$

The NR-NQ limit $(k_B T \ll mc^2)$ for the asymptotic expressions for $\beta_R, u_R - mc^2$ and $s_R$ are

$$\begin{cases} \beta_R \cong \dfrac{1}{2 mc^2 k_B T} = \dfrac{\beta}{2 mc^2} \\ u_R - mc^2 \cong \dfrac{3}{2} k_B T \\ s_R \cong \dfrac{3}{2} k_B + k_B ln[\dfrac{V}{h^3} \left(\dfrac{2\pi m}{\beta}\right)^{3/2}] \end{cases} \tag{51}$$

The approximate expressions of $u_R - mc^2$ and $s_R$ in (51) correspond to the classical (NQ-NR) expressions of the internal energy and entropy that can be deduced form the relations in (3) for $N = 1$ (single particle).

From the expression of $\beta_R$ for non-relativistic limit in (51), In NR-Q limit, from (47) the asymptotic expressions of $u_{RQ} - mc^2$ and $s_{RQ}$ are:

$$\begin{cases} u_{RQ} - mc^2 \cong \dfrac{1}{2} k_B T \left[ Tr\left( \dfrac{\beta}{m}[\mathcal{B}] \, coth\left( \dfrac{\beta}{m}[\mathcal{B}] \right) \right) \right] \\ s_{RQ} \cong [\dfrac{k_B \beta}{2m} tr\left( [\mathcal{B}] \, coth\left( \dfrac{\beta}{m}[\mathcal{B}] \right) \right) - k_B \ln\left\{ det\left[ sh\left( \dfrac{\beta}{m}[\mathcal{B}] \right) \right] \right\} \end{cases} \quad (52)$$

These asymptotic of $u_{RQ} - mc^2$ and $s_{RQ}$ in (52) correspond to the NR-Q expressions of the internal energy and entropy that can be deduced from the relations in (32) for $N = 1$.

### 4.2. Relativistic quantum partition function of the ideal gas

As in the section 3.2, we consider an ideal gas which is a set of $N$ indiscernible particles. However, we will firstly consider the calculation of the relativistic quantum partition function of a set of $N$ discernibles particles and we will deduce the partition function of indiscernible particles after. Given the relations (23) and (36), we may suppose that the relativistic Hamiltonian operator $\boldsymbol{H}$ and density operator $\boldsymbol{\rho}$ of a system composed by $N$ discernible particles at equilibrium with a heat bath are given (in their "mean rest frame") by the relation

$$(\boldsymbol{H})^2 = (E_0)^2 + 2 \sum_{k=1}^{N} \sum_{l=1}^{3} (2\boldsymbol{z}_l^{k\dagger} \boldsymbol{z}_l^k + 1) \mathcal{B}_{ll} c^2 \quad (53)$$

$$\boldsymbol{\rho} = e^{-\beta_R (\boldsymbol{H})^2} \quad (54)$$

in which:

- $E_0$ is the rest energy of the system. If it is supposed that the $N$ particles have the same mass $m$, one has the relation $E_0 = Nmc^2$.
- The operators $\boldsymbol{z}_l^k$ and $\boldsymbol{z}_l^{k\dagger}$ are the analogous of the operators $\boldsymbol{z}_l$ and $\boldsymbol{z}_l^\dagger$ in the relation (23) for the $k^{th}$ particle.
- The $\mathcal{B}_{ll}$ are the variances of the components of the linear momentum: we may suppose that they are the same for all atoms at the thermodynamic equilibrium.
- $\beta_R$ is a factor to be determined. It is not necessarily the same factor as in the single-particle density operator (36). More precisely, its expression is found in the relation (50).

The eigenvalues of the operator $(\boldsymbol{H})^2$ are of the form (it is supposed that the particles have the same mass $m$)

$$(\varepsilon_n)^2 = N^2 m^2 c^4 + 2 \sum_{k=1}^{N} \sum_{l=1}^{3} (2n_l^k + 1) \mathcal{B}_{ll} c^2 \quad (55)$$

The index $n$ in $\varepsilon_n$ refers to the $3N$-uplet of positive integers numbers $(n_1^1, n_2^1, n_3^1, \ldots, n_1^N, n_2^N, n_3^N)$. It can be deduced from the relations (54) and (55) that the expression of the partition of the particles is

$$Z = Tr\left( e^{-\beta_R (\boldsymbol{H})^2} \right) = \frac{e^{-\beta_R N^2 m^2 c^4}}{[8 sh(2\beta_R c^2 \mathcal{B}_{11}) sh(2\beta_R c^2 \mathcal{B}_{22}) sh(2\beta_R c^2 \mathcal{B}_{33})]^N} \quad (56a)$$

$$= \frac{e^{-\beta_R N^2 m^2 c^4}}{[8 det\{sh(2\beta_R c^2 [\mathcal{B}])\}]^N} \quad (56b)$$

If we suppose now that the particle are undiscernible, an approximate expression (which do not take into account the nature of the particles: bosons or fermions) of the relativistic quantum partition function $\mathcal{Z}_{RQ}$ which take into account the indiscernibility that can be deduced from relations $(56a)$ and $(56b)$ is

$$\mathcal{Z}_{RQ} = \frac{\mathcal{Z}}{N!} = \frac{e^{-\beta_R N^2 m^2 c^4}}{N! \, [8 det\{sh(2\beta_R c^2 [\mathcal{B}])\}]^N} \tag{57}$$

### 4.3. Relativistic corrections to the ideal gas model

We may found the non-quantum (but relativistic) approximate expression $\mathcal{Z}_R$ of the partition function $\mathcal{Z}_{RQ}$ by using an approximation analogous to what were used in the relations (28) and (40). The expression that is found is

$$\mathcal{Z}_R = \frac{V^N}{N! \, h^{3N} c^{3N}} \left(\frac{\pi}{\beta_R}\right)^{\frac{3N}{2}} e^{-\beta_R N^2 m^2 c^4} \tag{58}$$

$\mathcal{Z}_R$ is the R-NQ partition function of the ideal gas. Any thermodynamics properties can be deduced from it. The internal energy $U_R$ and entropy $S_R$ can, for instance, be found with relations analogous to the relations in (48)

$$\begin{cases} (U_R)^2 = -\frac{\partial[ln\mathcal{Z}_R]}{\partial \beta_R} = N^2 m^2 c^4 + \frac{3N}{2\beta_R} \\ S_R = k_B[ln\mathcal{Z}_R + \beta_R (U_R)^2] = ln[\frac{V^N}{N! \, h^{3N} c^{3N}} \left(\frac{2\pi[U^2 - N^2 m^2 c^4]}{3}\right)^{3N/2}] + \frac{3N}{2} k_B \end{cases} \tag{59}$$

From thermodynamic, we have a relation analogous to the relation (49), one has then

$$\left(\frac{\partial S_R}{\partial U_R}\right)_V = \frac{1}{T} \Leftrightarrow \frac{3N k_B U_R}{(U_R)^2 - m^2 c^4} = \frac{1}{T} \tag{60}$$

The explicit expressions of $\beta_R$, $U_R$ and $S_R$ can be deduced from the relations (59) and (60). One obtains

$$\begin{cases} \beta_R = \frac{1}{2 k_B T U_R} = \frac{3 k_B T - \sqrt{9(k_B T)^2 + 4 m^2 c^4}}{4 N m^2 c^4 k_B T} \\ U_R = \frac{1}{2 k_B T \beta_R} = N \frac{3 k_B T + \sqrt{9(k_B T)^2 + 4 m^2 c^4}}{2} \\ S_R = \frac{3}{2} N k_B + ln[\frac{V^N}{N! \, h^{3N} c^{3N}} \left(\frac{\pi}{\beta_R}\right)^{\frac{3N}{2}}] \end{cases} \tag{61}$$

The expressions in the relation (50) are obtained from (61) for $N = 1$. It may be verified that one obtains in the non-relativistic limit ($k_B T \ll mc^2$) for the asymptotic expressions for $\beta_R, U_R - N m c^2$ and $S_R$:

$$\begin{cases} \beta_R \cong \dfrac{1}{2Nmc^2 k_B T} = \dfrac{\beta}{2Nmc^2} \\ U_R - mc^2 \cong \dfrac{3}{2} N k_B T \\ S_R \cong \dfrac{5}{2} k_B + k_B ln[\dfrac{V}{h^3} \left(\dfrac{2\pi m}{\beta}\right)^{3/2}] \end{cases} \qquad (62)$$

The approximate expressions of $U_R - mc^2$ and $S_R$ in (62) correspond to the classical (non-quantum and non-relativistic) expressions of the internal energy and entropy corresponding to the relations in (3) (the Stirling's approximation was also used for the entropy). One can deduce from the relation (61) the expression of the Relativistic Free energy $F_R = U_R - TS_R$ and the pressure $P = -\left(\dfrac{\partial F_R}{\partial V}\right)_{T,N}$. The thermodynamic equation of state of the relativistic ideal gas can be deduced from the expression of the pressure and this equation is exactly similar to the classical equation of state (4). It means that the relativistic corrections did not bring change to the thermodynamic equation of state.

### 4.4. Relativistic Quantum corrections to the ideal gas model

The relativistic quantum partition function of the ideal gas is given by the relation (57). The explicit expression of the factor $\beta_R$ corresponding to this relation (57) is given in the relation (61). The relations between the elements $\mathcal{B}_{11}, \mathcal{B}_{22}$ and $\mathcal{B}_{33}$ of the momentum variance matrix $[\mathcal{B}]$ with $\beta_R$ and the characteristic of the volume containing the gas can be deduced as in the establishment of the relations in (41) by using the fact that in the R-NQ limit one has

$$\mathcal{Z}_{RQ} \cong \dfrac{e^{-\beta_R N^2 m^2 c^4}}{N! \, (64)^N (\beta_R)^{3N} c^{6N} \mathcal{B}_{11} \mathcal{B}_{22} \mathcal{B}_{33}} = \mathcal{Z}_R = \dfrac{V^N}{N! \, h^{3N} c^{3N}} \left(\dfrac{\pi}{\beta_R}\right)^{\frac{3N}{2}} e^{-\beta_R N^2 m^2 c^4} \qquad (63)$$

We obtain for instance for the case of a parallelepipedic volume $V$ equal to $L_1 L_2 L_3$

$$\begin{cases} \mathcal{B}_{11} = \dfrac{h}{4cL_1 \sqrt{\pi \beta_R}} = \dfrac{hcm}{2L_1} \left(\dfrac{N k_B T}{\pi(3k_B T - \sqrt{9(k_B T)^2 + 4m^2 c^4})}\right)^{1/2} \\ \mathcal{B}_{22} = \dfrac{h}{4cL_2 \sqrt{\pi \beta_R}} = \dfrac{hcm}{2L_2} \left(\dfrac{N k_B T}{\pi(3k_B T - \sqrt{9(k_B T)^2 + 4m^2 c^4})}\right)^{1/2} \\ \mathcal{B}_{33} = \dfrac{h}{4cL_3 \sqrt{\pi \beta_R}} = \dfrac{hcm}{2L_3} \left(\dfrac{N k_B T}{\pi(3k_B T - \sqrt{9(k_B T)^2 + 4m^2 c^4})}\right)^{1/2} \end{cases} \qquad (64)$$

From (57), one can deduce the expressions of the RQ thermodynamic properties. The internal energy $U_{RQ}$ and entropy $S_{RQ}$ can be for instance deduced using relations that are analogous to the relations (45) and (46). One obtains the relations

$$\begin{cases} (U_{RQ})^2 = -\dfrac{\partial [ln\mathcal{Z}_{RQ}]}{\partial \beta_R} \\ S_{RQ} = k_B \left[ ln(\mathcal{Z}_{RQ}) + \beta_R (U_{RQ})^2 \right] \end{cases} \qquad (65)$$

Using the relation (57), one can deduce the following explicit expressions

$$\begin{cases} U_{RQ} = N\sqrt{\{m^2c^4 + tr(c^2[\mathcal{B}]\coth(2\beta_R c^2[\mathcal{B}]))\}} \\ S_{RQ} = N^2 k_B \beta_R tr(c^2[\mathcal{B}]\coth(2\beta_R c^2[\mathcal{B}])) - k_B \ln\{N! [8det\{sh(2\beta_R c^2[\mathcal{B}])\}]^N\} \end{cases} \quad (66)$$

For $N = 1$ (single particle), the expressions in the relation (47) are recovered from (64). In the relativistic but non-quantum limit $(2\beta_R c^2 \mathcal{B}_{ll}) \ll 1$, one can deduce, as approximation from (64) the expressions of $U_R$ and $S_R$ in (61). And in the quantum but non-relativistic limit $(k_B T \ll mc^2)$ one can deduce, as approximation of (64) the expressions of $U_Q + Nmc^2$ and $S_Q$ corresponding to the relations (32).

From the relation in (64), one can also deduce the free energy $F_{RQ} = U_{RQ} - TS_{RQ}$. For the pressure, like in the R-NQ case, we do not have a scalar like in the classical case but a matrix $[P]$ which is linked to the matrix $[\mathcal{B}]$. One obtains, for instance, for a parallelepipedic volume $V$ equal to $L_1 L_2 L_3$

$$\begin{cases} P_{11} = \dfrac{Nk_B T}{V}[2\beta_R c^2 \mathcal{B}_{11} \coth(2\beta_R c^2 \mathcal{B}_{11})] + \dfrac{N^2 \mathcal{B}_{11} c^2}{V}\left(\dfrac{U_{RQ}-U_R}{U_{RQ} U_R}\right)\dfrac{sh(4\beta_R c^2 \mathcal{B}_{11}) - 4\beta_R c^2 \mathcal{B}_{11}}{4[sh(2\beta_R c^2 \mathcal{B}_{11})]^2} \\ P_{22} = \dfrac{Nk_B T}{V}[2\beta_R c^2 \mathcal{B}_{22} \coth(2\beta_R c^2 \mathcal{B}_{22})] + \dfrac{N^2 \mathcal{B}_{22} c^2}{V}\left(\dfrac{U_{RQ}-U_R}{U_{RQ} U_R}\right)\dfrac{sh(4\beta_R c^2 \mathcal{B}_{22}) - 4\beta_R c^2 \mathcal{B}_{22}}{4[sh(2\beta_R c^2 \mathcal{B}_{22})]^2} \\ P_{33} = \dfrac{Nk_B T}{V}[2\beta_R c^2 \mathcal{B}_{33} \coth(2\beta_R c^2 \mathcal{B}_{33})] + \dfrac{N^2 \mathcal{B}_{33} c^2}{V}\left(\dfrac{U_{RQ}-U_R}{U_{RQ} U_R}\right)\dfrac{sh(4\beta_R c^2 \mathcal{B}_{33}) - 4\beta_R c^2 \mathcal{B}_{33}}{4[sh(2\beta_R c^2 \mathcal{B}_{33})]^2} \end{cases} \quad (67)$$

The thermodynamic equation of state can be deduced easily from the expression of the elements of the pressure matrix in the relation (67). In the R-NQ limit $(2\beta_R c^2 \mathcal{B}_{11} \ll 1)$ or in the classical (NR-NQ) limit this equation of state is reduced to the classical equation (4): as already seen in the previous section, the thermodynamic equation of state is the same for relativistic and classical cases. In the NR-Q limit $(k_B T \ll mc^2)$, the relations in (67) give as asymptotic limit the relations in (34).

## 5. Results and discussions

Following the content of the previous sections, we have obtained the expressions of the partition functions, the internal energy, the entropy, the pressure and the thermodynamic equation state for the RQ case, the R-NQ case and the NR-Q case:

- The RQ case corresponds to the most general corrections that can be deduced from the approach that was considered. The expressions of the ideal gas partition function for this case is given by the relation (57). The explicit expression of the factor $\beta_R$ which appears in this relation (57) can be found in the relation (61). The expressions of the internal energy, the entropy are given in the relations (66). The pressure is a matrix (a tensor) and the expressions of its components are given, with the example of a parralelepipedic volume, in the relation (67). The Thermodynamic equation of state can be deduced easily form the expressions of the pressure matrix elements given by the relation (67).

- The R-NQ case corresponds to the asymptotic limit of the RQ case when quantum effects can be neglected. It corresponds to the conditions $2\beta_R c^2 \mathcal{B}_{ll} \ll 1$ and $k_B T \approx mc^2$. More explicitly, given the expression of the momentum variances $\mathcal{B}_{ll}$ in the relation (64), these conditions correspond to large volume and/or very high temperature. The R-NQ expression of the internal energy and entropy are given in the relation (61). It can be deduced that the expression of the pressure and the Thermodynamic equation of state is the same as in the NR-NQ (i.e. classical) case. The relativistic corrections did not affect the expressions of the pressure and the equation of state.

- The NR-Q case corresponds to the limit of the RQ case when relativistic effects can be neglected. It corresponds to the condition $k_B T \ll mc^2$. The corresponding expressions of the internal energy and entropy are given in the relation (32). The expression of the pressure, which is a matrix, is given in the relation (33), the explicit example of the parralepipedic volume corresponds to the relation (34). The thermodynamic equation of state which can be deduced easily from the relation (33) is given in the relation (35). The obtained results show that the quantum effects are particularly significant for confined volume and low temperature.

In the NR-NQ i.e. classical limit which corresponds to $2\beta_R c^2 \mathcal{B}_{ll} \ll 1$ and $k_B T \ll mc^2$, the classical properties of the Maxwell-Boltzmann ideal gas corresponding to the relations (2), (3) and (4) are obtained.

Apart from the use of an explicit quantum phase space, a main difference between our approach and the standard formalism of statistical mechanic as considered for instance in the references [2-4] is the introduction of a new distribution for the relativistic case. In fact the relativistic relations (36) and (54), for instance, does not correspond to the standard canonical distribution or to another known distribution. There is, in particular, significant differences between them and other known relativistic distributions like those considered in the references [9, 11-12, 35-36]. A main characteristic of these new distributions is the introduction of the factor $\beta_R$ which is given in the relation (50) for a single particle and in the relation (61) for many particles.

The fact that the pressure becomes a matrix for the RQ and NR-Q cases corresponds to the existence of quantum size and shape effects for confined volume.

## 6. Conclusion

Partition functions and thermodynamic properties for quantum and relativistic ideal gas have been deduced by the use of Hamiltonian operators which are compatible with the concept of quantum phase space that was considered in the references [22-24]. The new quantum and relativistic corrections are revealed to be particularly useful to describe the deviation from classical behavior of a gas at low temperature and in confined volume for the quantum case and when the thermal energy is comparable to the rest energy for the relativistic case. The properties of classical ideal gas are obtained as asymptotic limits. This work shows that the formulation of statistical physics and thermodynamics can be based on the concept of quantum phase space.

Our approach has permitted to obtain new kinds of relativistic and quantum corrections for the Maxwell-Boltzmann ideal gas model. We choose the qualification "Maxwell-Boltzmann ideal gas" because the quantum nature of the particles (bosonic or fermionic) were not considered. The obtained results are then in particular limited by this approximation which correspond to the approximates expressions of the N-particles partition functions corresponding to the relations (30) for the NR-Q case and the relation (57) for the RQ case. However, the results thus obtained may have some importance and it is possible to consider the extension of the approach to the cases of Bose gas and Fermi gas. These obtained results may also help, in particular in the study and resolutions of some issues and problems related to relativistic thermodynamics like those that were considered in the references [8, 10, 37-42].

# References

1. Wigner, E. P.: On the quantum correction for thermodynamic equilibrium. Phy. Rev. 40 749-759 (1932)
2. Hardi, R. J., Binek, C.: Thermodynamics and statistical mechanics, Wiley (2014)
3. Reif, F.: Fundamental of statistical and thermal physics, Waveland Press (2009)
4. Guénault, T.: Statistical physics, Springer Dordrecht (1995).
5. Attard, P.: Quantum Statistical Mechanics, IOP Publishing (2015)
6. Kosloff, R.: Quantum Thermodynamic : A Dynamical Viewpoint, Entropy 15, 2100-2128 (2013)
7. Raoelina Andriambololona, "Mécanique quantique", Collection LIRA, INSTN-Madagascar, 1990
8. Planck, M.: Zur Dynamik bewegter systeme, Annalen der physic, 331 (6) : 1-34 (1908)
9. Jüttner, F.: Das MaxwellscheGesetz der Geschwindigkeitsverteilung in der Relativtheorie". Annalen der Physik. **339** (5): 856–882 (1911)
10. Ott, H.: Lorentz-transformation der Warme und der Temperatur, Zeitschrift fur Physik 175, 70-104 (1963)
11. Rovelli, C. : General relativistic statistical mechanics, Physical Review D 87, 0845055 (2013)
12. Becattini, F. : Covariant Statistical Mechanics and the Stress-Energy Tensor, Physical Review Letters 108, 244502 (2012)
13. Vinjanampathy, S., Anders, J.: Quantum thermodynamics, Contemporary Physics57(4) 545–579 (2016)
14. Deffner, S., Campbell, S. : Quantum Thermodynamics, Morgan & Claypool Publishers (2019)
15. Mahler, G.: Quantum Thermodynamic Processes, CRC Press, Taylor & Francis Group (2015)
16. Aydin, A., Sisman A.: Dimensional transitions in thermodynamic properties of ideal Maxwell–Boltzmann gases, PhysicaScripta 90 045208 (2015)
17. Ozturk, Z.F., Sisman A.: Quantum size effects on the thermal and potential conductivities of ideal gases-PhysicaScripta.80(6) 654-662 (2009)
18. Aydin, A., Sisman A.: Quantum shape effects and novel thermodynamic behaviors at nanoscale, Physics Letter A. 383(7)655-665 (2019)
19. Sisman A.: Surface dependency in the thermodynamics of ideal gases. Journal of Physics A: Math. Gen. 37 (43) 11353-11361 (2004)
20. Aydin, A., Sisman A.: Quantum oscillations in confined and degenerate Fermi gases. I. Half-vicinity model.PhysicsLetter A. 382(27) 1807-1812 (2018)
21. Pang, H.: The pressure exerted by a confined ideal gas. Journal of Physics A: Math. Theor. 44 365001 (2011)
22. Ranaivoson, R.T., Raoelina Andriambololona, Hanitriarivo, R., Raboanar R. : Linear Canonical Transformations in relativistic quantum physics Physica Scripta 96 (6) 065204 (2021)
23. Ranaivoson, R.T., Raoelina Andriambololona, Hanitriarivo, R., Ravelonjato, R. H. M.: Invariant quadratic operators associated with Linear Canonical Transformations and their eigenstates. Journal of Physics Communication. 6 095010 (2022)
24. Ranaivoson, R.T., Hejesoa, V.S., Raoelina Andriambololona, Rasolofoson, N.G., Rakotoson, H., Rabesahala, J., Rarivomanantsoa, L., Rabesiranana, N.: Highlighting relations between Wave-particle duality, Uncertainty principle, Phase space and Microstates, arXiv:2205.08538 [quant-ph], (2022)
25. Quarati, P, Lissia, M.: The Phase Space Elementary Cell in Classical and Generalized Statistics, Entropy 15(10), 4319-4433 (2013).
26. Priyank Shah.:A multiobjective thermodynamic optimization of a nanoscalestrirling engine operated with Maxwell-Boltzmann gas, Heat Transfer. Asian Research 1-20 (2019)
27. Edward L, Wolf. : Nanophysics and Nanotechnology: an introduction to modern concepts in Nanoscience. WILEY-VCH Verlag GmbH & Co. KGaAWeinheim (2006)
28. Signe Kjelstrup et al.:Bridging scales with thermodynamics: from nano to macro. Advancez in Natural Sciences: Nanoscience and Nanotechnology. 5 0230022014 (2014.)
29. RaoelinaAndriambololona.: Algèbre Linéaire et Multilinéaire et Applications, 3 Tome, Collection LIRA, INSTN-Madagascar (1985)
30. Curtright, T.L., Zachos, C.K.: Quantum Mechanics in Phase Space, arXiv:1104.5269v2 [physics.hist-ph]", Asia Pacific Physics Newsletter, V1, Iss 1, pp 37-46, May 2012
31. Rundle, R. P., Everit, M. J.: Overview of the phase space formulation of quantum mechanics with application to quantum technologies, Adv. Quantum Technol. 4 (6) 2100016, (2021)
32. Weyl, H.: Quantenmechanik und Gruppentheorie. ZeitschriftfürPhysik (in German). 46 (1–2): 1–46, (1927)
33. Groenewold, H.J.: On the Principles of elementary quantum mechanics, Physica 12, (1946)
34. Moyal, J.E.: Quantum mechanics as a statistical theory, Proceedings of the Cambridge Philosophical Society 45, 99–124, (1949)

35. Chacon-Acosta, G., Dagdug, L. : Manifestly covariant Jüttner distribution and equipartition theorem, Physical Review E 81, 021126 (2010)
36. Zaninetti, L.: New Probability Distributions in Astrophysics: IV. The Relativistic Maxwell-Boltzmann Distribution, International Journal of Astronomy and Astrophysics, 10, 302-313 (2020).
37. C. Farías, V. A. Pinto, P. S. Moya, What is the temperature of a moving body?. Sci Rep 7, 17657 (2017)
38. Landsberg, P.T.: Does a moving body appear cool? , Nature 214, 903-904 (1967)
39. Landsberg, P.T.: Laying the ghost of relativistic temperature transformation, Physics letter A 223, 401-403 (1996)
40. Landsberg, P.T.: the impossibility of a universal relativistic temperature transformation, Physica A 340, 92–94 (2004)
41. Sewell, G.L.: On the question of temperature transformations under Lorentz and Galilei boosts, J. Phys. A: Math. Theor. 41 382003 (2008)
42. Papadatos, N.: Relativistic quantum thermodynamics of moving systems, Physical review D 102, 085005 (2020)